\documentclass{article}

\PassOptionsToPackage{numbers,compress}{natbib}
\usepackage[preprint]{neurips_2025}
\usepackage{graphicx}
\usepackage{float}
\usepackage{subcaption}
\usepackage{amsmath}
\usepackage{hyperref}
\usepackage{enumitem}
\usepackage[skip=0pt, font=small]{subcaption}

\usepackage[utf8]{inputenc} 
\usepackage[T1]{fontenc}    
\usepackage{hyperref}       
\usepackage{url}            
\usepackage{booktabs}       
\usepackage{amsfonts}       
\usepackage{nicefrac}       
\usepackage{microtype}      
\usepackage{xcolor}         

\title{Double Descent and Overparameterization in\\Particle Physics Data}

\author{
  Matthias Vigl\\
  Technical University of Munich\\
  \texttt{matthias.vigl@tum.de} \\
   \And
   Lukas Heinrich \\
   Technical University of Munich \\
 \texttt{lukas.heinrich@tum.de} \\
}

\begin{document}

\maketitle

\begin{abstract}
  Recently, the benefit of heavily overparameterized models has been observed in machine learning tasks: models with enough capacity to easily cross the \emph{interpolation threshold} improve in generalization error compared to the classical bias-variance tradeoff regime. We demonstrate this behavior for the first time in particle physics data and explore when and where `double descent' appears and under which circumstances overparameterization results in a performance gain.
\end{abstract}

\section{Introduction}
Particle physics aims to understand the smallest constituents of matter and their interactions and to search for new fundamental particles to explain unsolved phenomena such as the nature of dark matter. Due to the high-dimensional, heterogeneous, and multimodal data particle detectors produce, the computational analysis of the data relies on increasingly large neural network components for a wide range of tasks, such as jet-tagging~\cite{Kasieczka_2019,Komiske:2018cqr}, particle reconstruction~\cite{Pata_2021,Di_Bello_2023}, detector simulation~\cite{Paganini_2018,Hashemi_2024} and many more. Recently, with the advent of foundation models~\cite{Leigh:2024ked, Golling:2024abg,Vigl:2024lat,Birk:2025wai,Birk:2024knn,Giroux:2025elr}, the field explores larger-scale, higher capacity models in order to push towards ultimate experimental sensitivity.

A key question in scaling neural networks is to understand their generalization behavior. Perhaps most strikingly, it has been observed that heavily overparameterized model can outperform models in the ``classical regime'' due to the ``double descent''~\cite{nakkiran2019deepdoubledescentbigger} phenomenon. Thanks to \emph{implicit bias} in the training dynamics~\cite{soudry2018implicit} the generalization of neural networks improves again after initially worsening due to heavy overfitting close to the interpolation regime, where the training loss first vanishes. The detailed behavior, however, is not universal but strongly depends on the nature of the chosen models and the application datasets. For example, whether or not the test-time performance in the overparameterized regime outperforms the best underparametrized model or whether the double-descent phenomenon persists under early-stopping model selection, requires domain-specific investigation. The phenomenon has been observed in the domains of natural images~\cite{stephenson2021epochwisedoubledescenthappens}, quantum systems~\cite{Kempkes:2025hiw} or protein folding~\cite{ahdritz2024openfold} but was not yet studied in particle physics data.

In this paper, we investigate generalization and overparameterization for particle physics data. We show that generalization behavior is indeed not uniform for the domain itself but is quite sensitive to task and dataset, motivating more detailed future experiments. We show for the first time explicit instances of epoch-wise and model-wise double descent. Additionally, we show that indeed instances can be found in data-constrained regimes, where overparameterized models outperform models in the classical regime. The experiments were performed on publicly available data, and we release code to facilitate future investigations of these phenomena in particle physics.

\section{Datasets and training setup}
\label{datasets}

We investigate overparameterization on typical tasks in particle physics using two representative public datasets from the ATLAS collaboration~\cite{PERF-2007-01}:

\begin{itemize}[leftmargin=0pt, label={}]
    \item[] \textbf{JetSet} This dataset~\cite{atlas2025jetset} consists of approximately 200M anti-$k_\text{T}$ R=0.4~\cite{Cacciari:2008gp} ``small-R'' jets from simulated top quark pair production. The dataset contains event-level, jet-level, track-level and truth hadron information used for training the GN2 jet flavor tagger~\cite{atlascollaboration2025transformingjetflavourtagging}. A detailed description of the dataset content can be found in the dedicated repository~\footnote{JetSet repository: \url{https://gitlab.cern.ch/atlas/open-data/transforming-jet-flavor/-/blob/main/vars_open.md}}
    \item[] \textbf{SUSY Wh1Lbb Channel Open Data Set} This dataset~\cite{atlas2024susy} consists of approximately 12M simulated events of signal chargino decay and 14 Standard Model (SM) background processes. The dataset contains object-level and event-level variables with systematic uncertainties.
\end{itemize}    

To disentangle the effect of model size on training dynamics and test set performance from the choice of optimizer parameters and learning rate schedulers, we use Adam optimizer~\cite{2014arXiv1412.6980K} with no weight decay across all experiments and model sizes, set a common batch size of 128 and $10^{-4}$ learning rate: the learning rate is kept constant when training multilayer perceptrons (MLPs), while a learning rate warmup for 5\% of the total straining steps is applied when training Transformer models.

\section{Jet $p_\mathrm{T}$ Regression: Model and Epoch-wise Double Descent}
\label{regression}
Particle jets emerge from cascading interactions in quantum chromodynamics (QCD) and produce a large number of collimated particles. For data analysis, these particles are clustered using agglomerative clustering~\cite{Cacciari:2008gp} into ``jets'' which are strongly related to the underlying generative process . Due to their high dimensionality, machine-learning approaches for particle jets are a major area of research in particle physics. Here, we demonstrate overparameterization behavior in the task of regressing the ``transverse momentum'', $p_T$, of the jet given the set of its constituent objects, i.e. the clustered tracks.

As a set-level prediction task, we choose a permutation equivariant two-layer transformer model, with model width $d_\mathrm{model}$ used as the scaling parameter. The transformed set of tokens is then pooled and the final $p_\mathrm{T}$ regression is computed using an MLP head network. The models are trained on 40k jets and we use mean-squared error (MSE) as the standard regression loss function. The loss is monitored for up to 4k epochs at increasing model width values ranging from $d_\mathrm{model} \in [4,2048]$. 

\begin{figure}[H]
    \centering

    \begin{subfigure}[b]{0.45\textwidth}
        \centering
        \includegraphics[width=\linewidth]{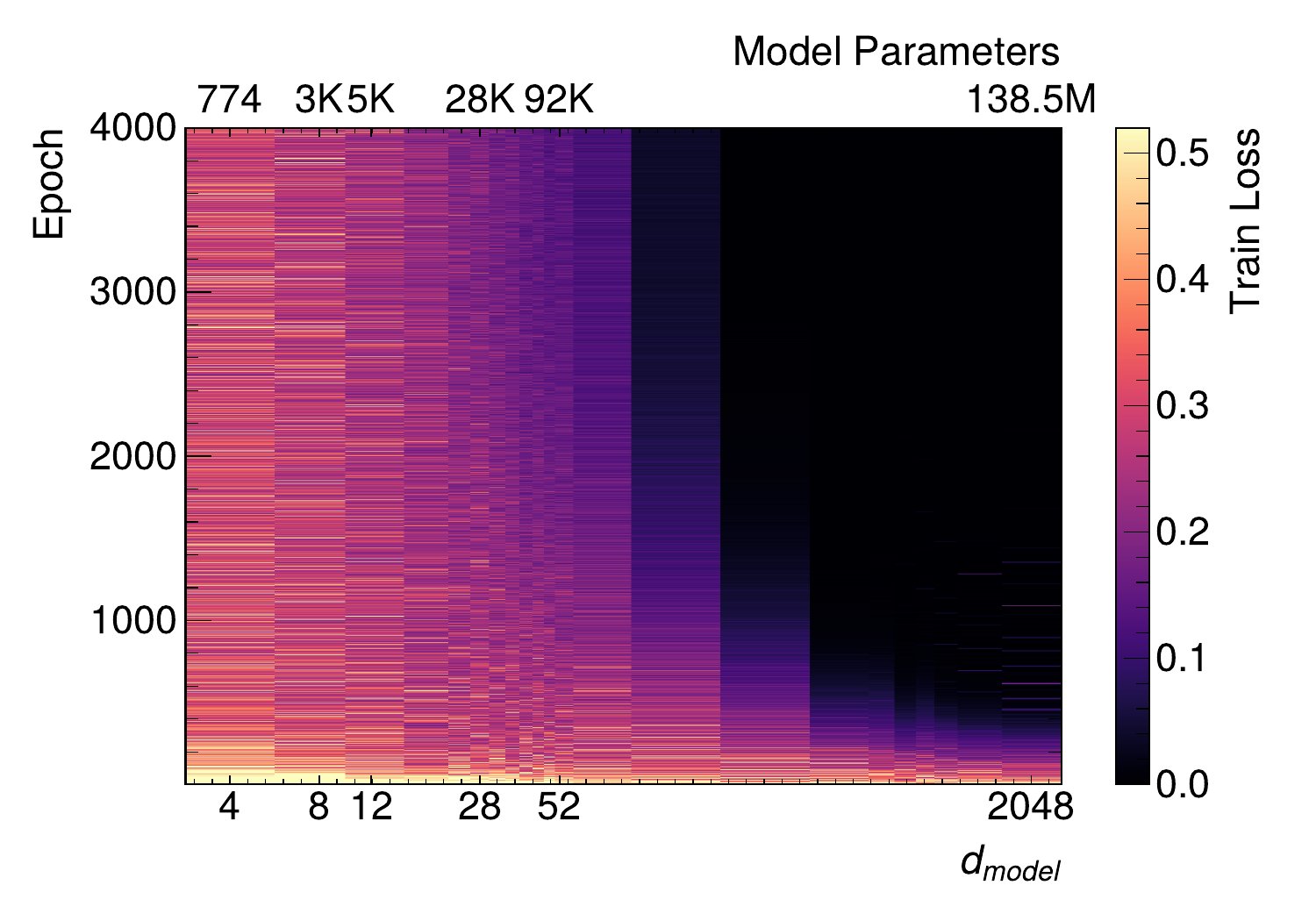}
        \caption{}
        \label{fig:pt_a}
    \end{subfigure}%
    \hspace{0.01\textwidth}%
    \begin{subfigure}[b]{0.45\textwidth}
        \centering
        \includegraphics[width=\linewidth]{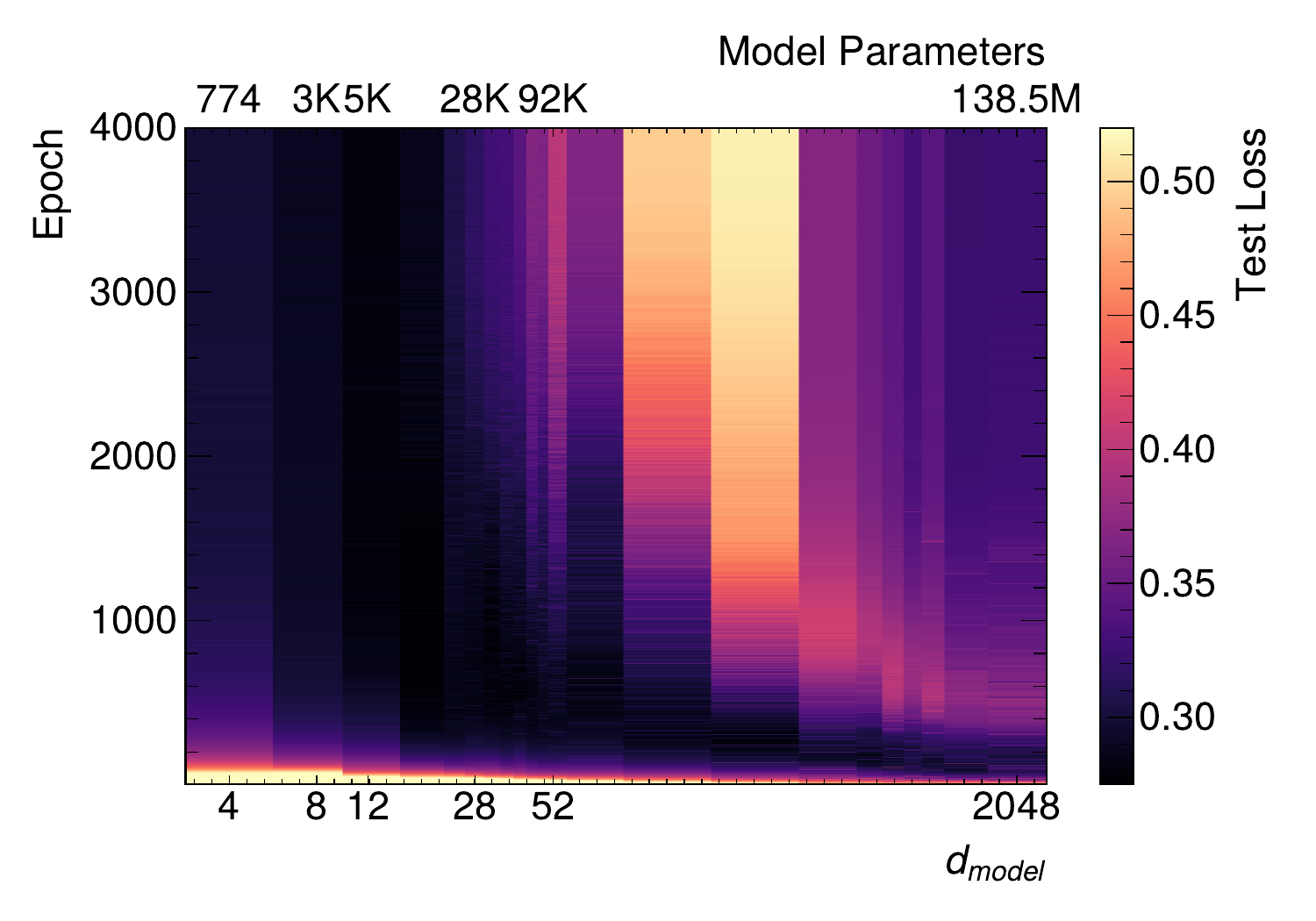}
        \caption{}
        \label{fig:pt_b}
    \end{subfigure}%

    \vspace{2mm} 

    \begin{subfigure}[b]{0.45\textwidth}
        \centering
        \includegraphics[width=\linewidth]{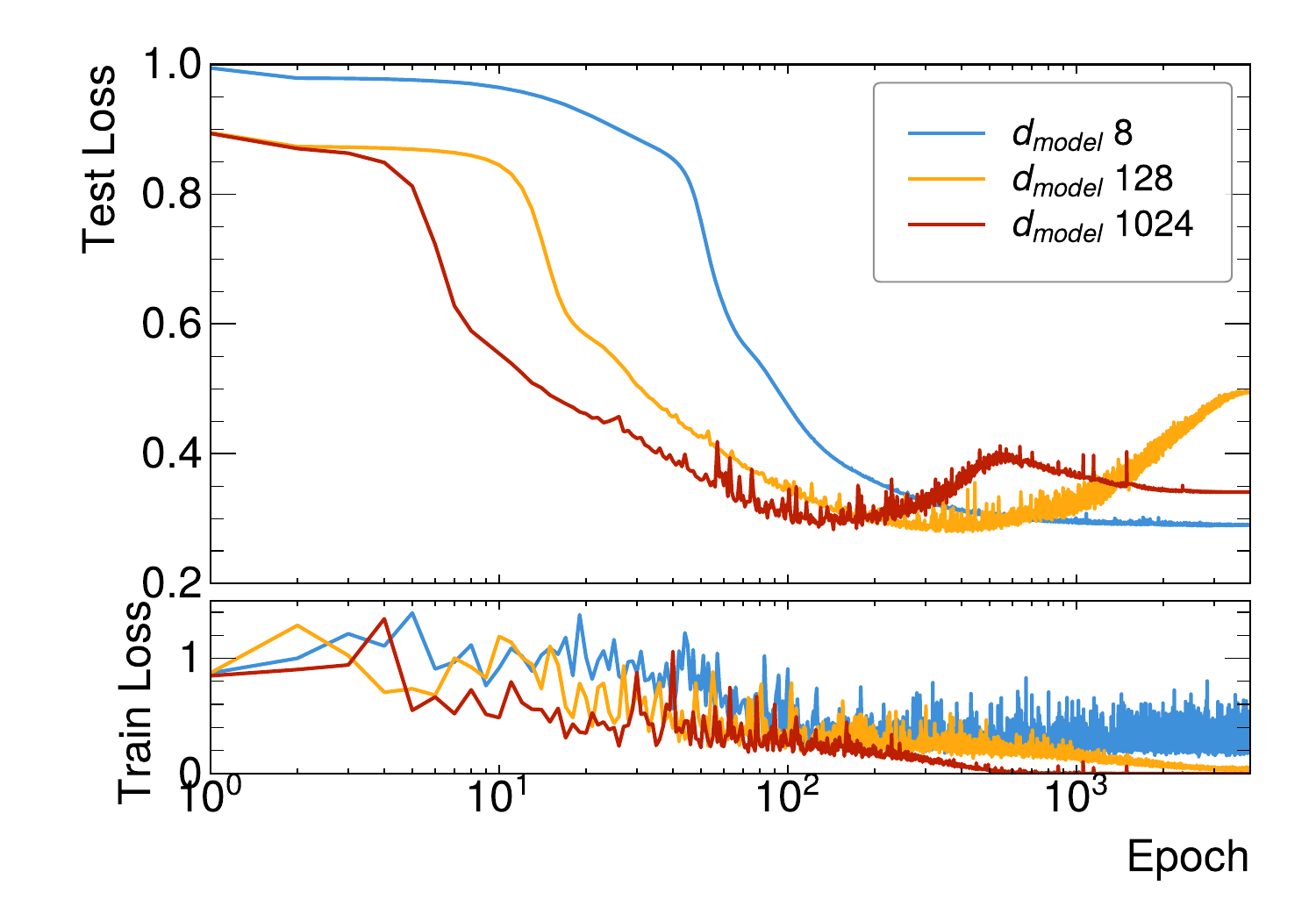}
        \caption{}
        \label{fig:pt_c}
    \end{subfigure}%
    \hspace{0.01\textwidth}%
    \begin{subfigure}[b]{0.45\textwidth}
        \centering
        \includegraphics[width=\linewidth]{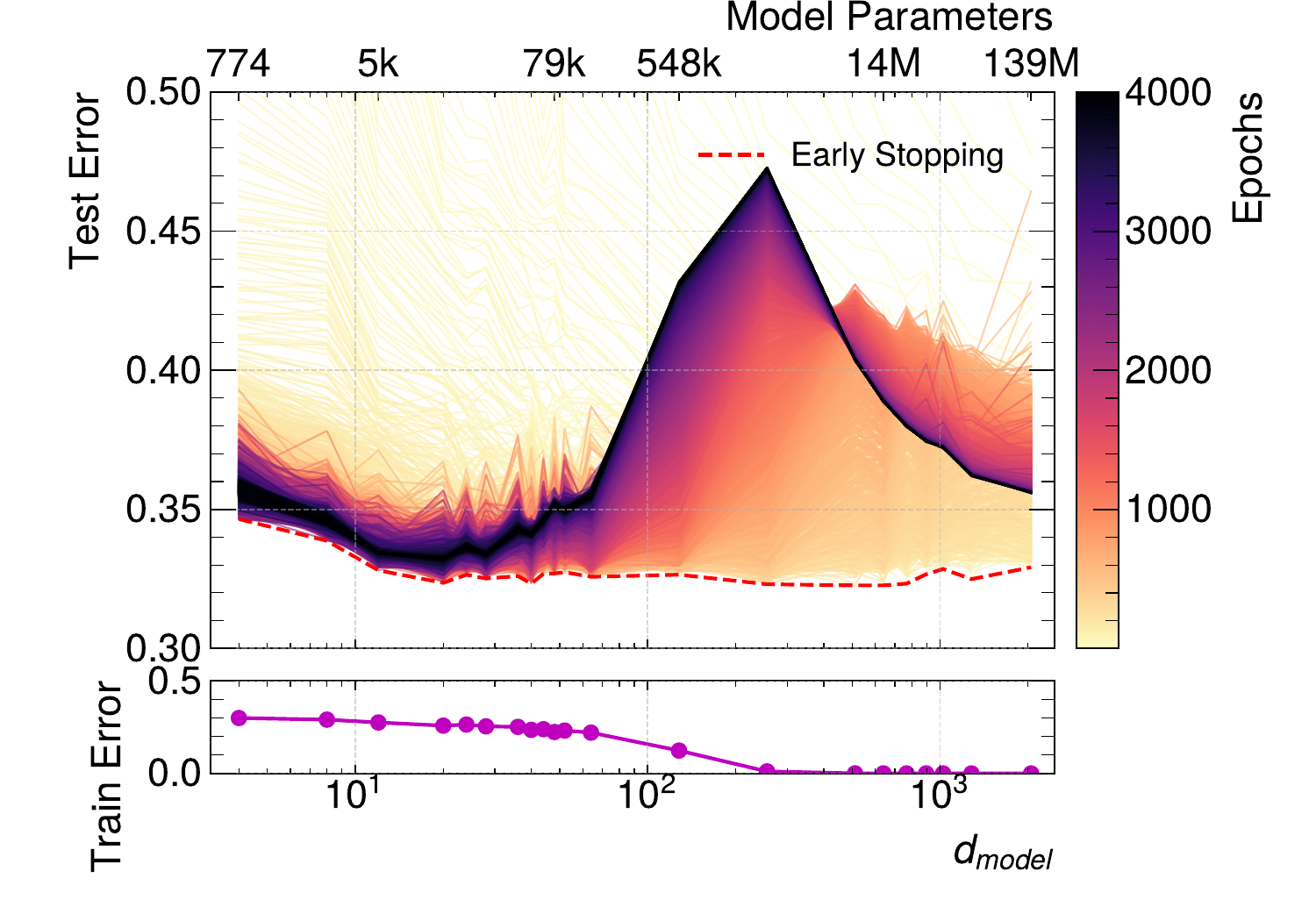}
        \caption{}
        \label{fig:pt_d}
    \end{subfigure}

    \caption{(a,b): Train and test jet $p_\mathrm{T}$ regression loss as a function of model size and epoch, model- and epoch-wise double descent can be observed looking at horizontal and vertical slices. (c): Vertical slices of plot (b) showing epoch-wise double descent. (d): Model-wise double descent.}
    \label{fig:pt}
\end{figure}

The achieved train and test loss are shown in Figures~\ref{fig:pt_a} and \ref{fig:pt_b} as a function of model capacity (i.e. model width) and the training epoch. Here, the typical features of the double descent phenomenon are for the first time visible in particle physics data. In the training loss, we can observe that for sufficiently large models the \emph{interpolation threshold} can be reached with sufficiently many training epochs. That is, the train loss vanishes and the models ``memorize'' the 40k training samples.

The test loss exhibits model-wise double descent as shown in Fig.~\ref{fig:pt_d} and is expected from generalization theory. In the model-wise direction the test loss is worst at the interpolation threshold, where the model capacity is just big enough to interpolate the data but does not have capacity for the implicit bias of the optimization procedure to select well-generalizing functions among the set of interpolating functions. For large models, the test-time performance decreases again.

Notably, the task also exhibits epoch-wise double descent, as shown in Fig.~\ref{fig:pt_c}. The test performance first improves over the course of training before the model starts overfitting but at sufficiently high epoch times, once the training data is ``memorized'' the generalization performance improves again.  The emergence of this phenomenon is theoretically less understood but is hypothesized to be related to differing speeds at which task-relevant features are learned~\cite{stephenson2021epochwisedoubledescenthappens}. Understanding which aspects of particle jet data contribute to this phenomenon is a interesting future research direction.

For both types of double descent, the second descent \emph{does not} outperform the test performance of the ``classical regime'', underlining again the dataset dependence of the benefit of overparameterization.

\section{Event classification: Prior Dependence and Early-Stopping Behavior}
\label{classification}

The rich context dependence of generalization performance of neural networks can be further investigated by comparing the behavior observed in jet $p_\mathrm{T}$ regression to that in high-level tasks such as event classification. In high energy physics event classification, the neural network is tasked with classifying a small set of high-level summary statistics extracted from high-dimensional multi-modal detector data with respect to their theoretical origin. A typical tasks is differentiating events likely to originate from known physics phenomena due to interactions described by the Standard Model of Particle Physics and those compatible with new phenomena such as Supersymmetry (SUSY)~\cite{MARTIN_1998} that could offer explanations for e.g. dark matter. For this task, we train a simple 3-layer MLP network to classify events from the SUSY dataset into multiple classes using a standard categorical cross-entropy loss. The scaling dimension in this case is the width of the hidden MLP layers.

\subsection{Prior Dependence of Epoch-wise Double Descent}

First, we investigate the nature of epoch-wise double descent more closely and can correlate the phenomenon empirically to prior choices. To this end, the network is tasked to classify summaries of collider events into one of four possible classes which represent the signal process (a supersymmetric particle model) and three major background processes (single top production, top-anti-top quark production and $W$-boson production associated with jets). As a posterior prediction task, the trained classifier depends strongly on the prior distribution chosen among the four classes, which we observe to also have an impact on generalization performance. 

\begin{figure}[H]
  \centering
    \includegraphics[width=0.44\linewidth]{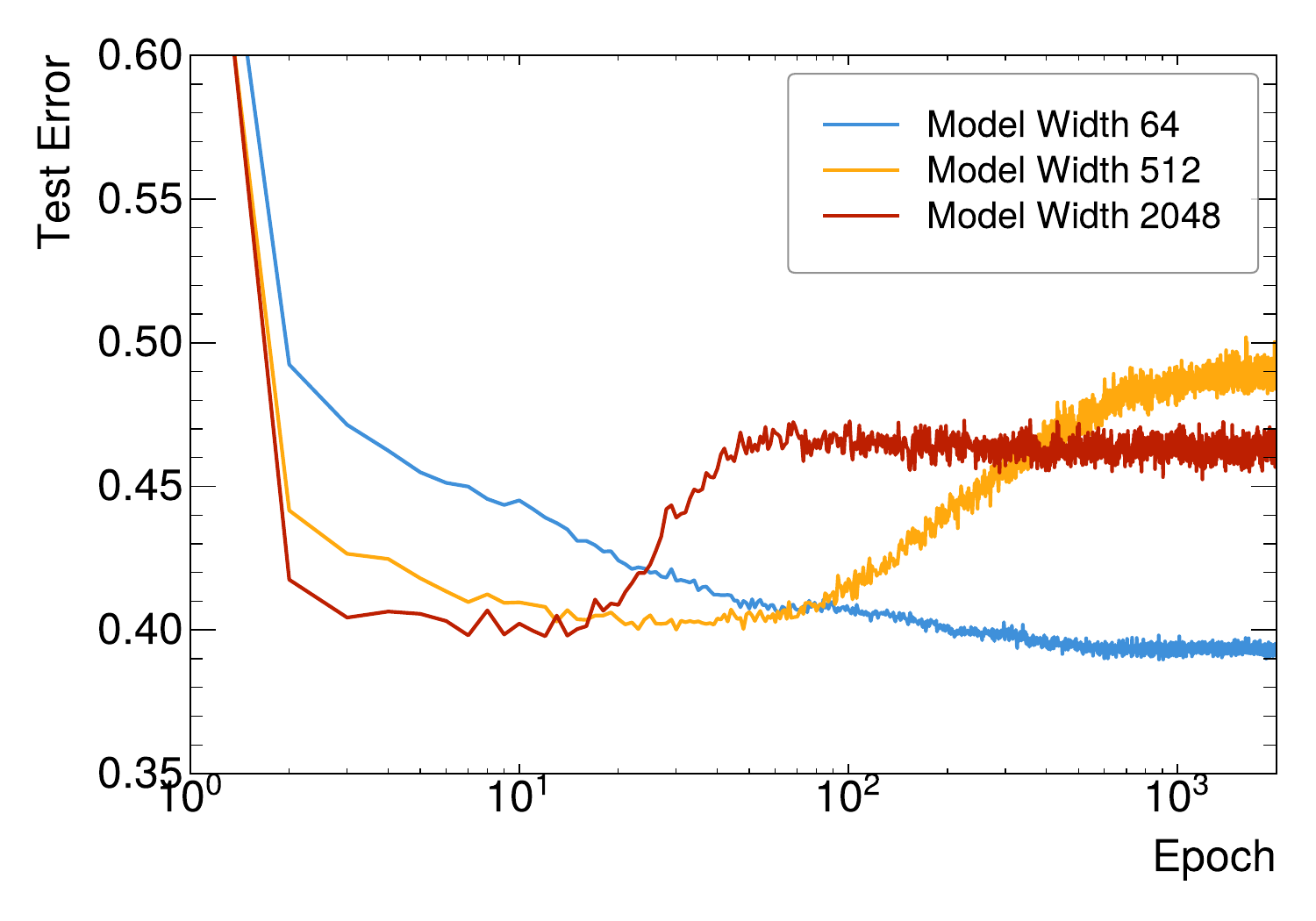}
    \includegraphics[width=0.44\linewidth]{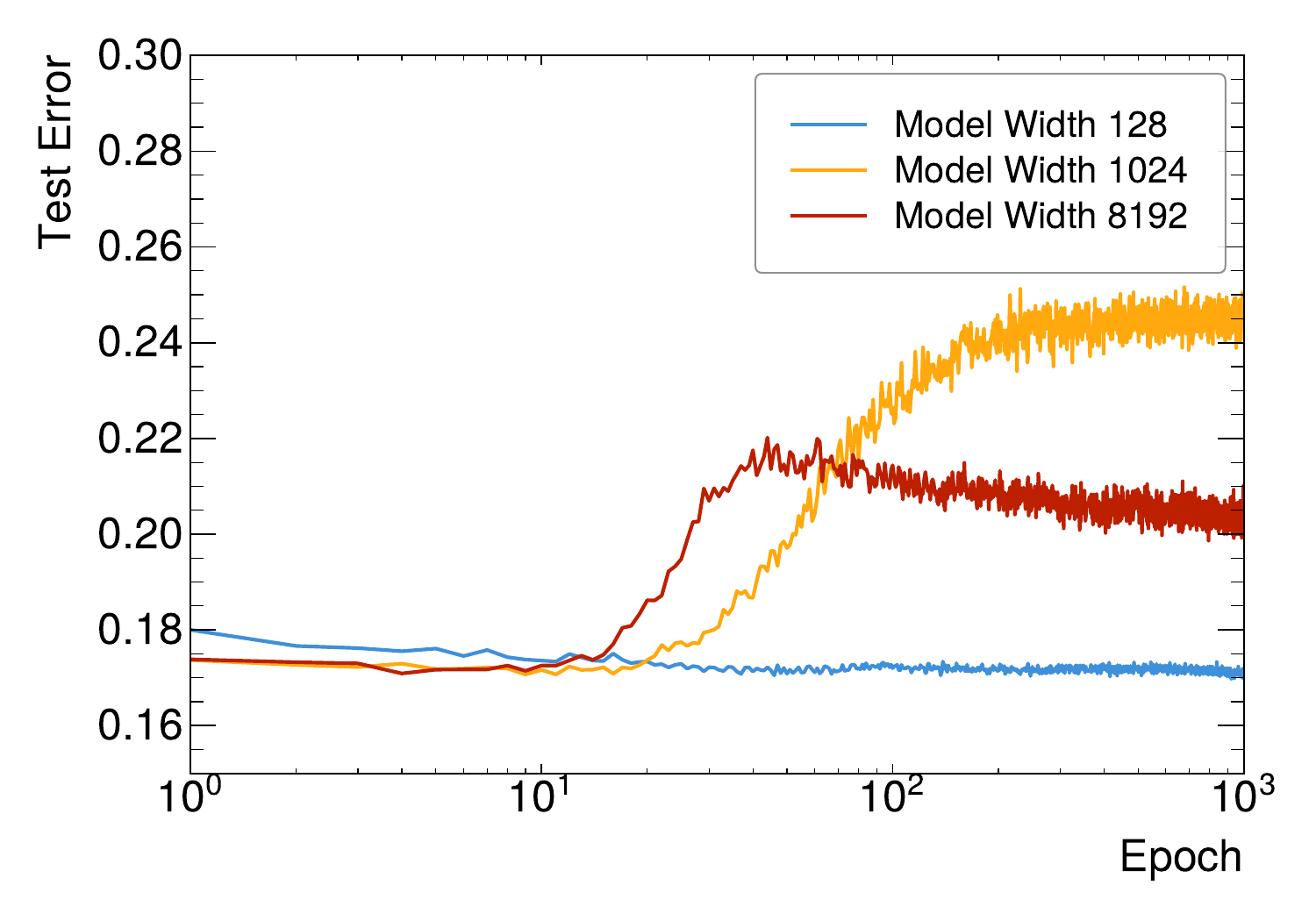}
  \caption{Large models trained on the 800k balanced dataset (left) do not exhibit epoch-wise double descent, whereas the same models trained on the 7M unbalanced dataset (right) do.}
  \label{fig:prior_dependence}
\end{figure}

In Fig.~\ref{fig:prior_dependence}, we compare test-loss trajectories for a balanced dataset of 800k events and an unbalanced dataset of 7M events, where processes are sampled proportionally to their occurrence in the full simulation, across the four target classes. We see that epoch-wise double descent is not a universal feature but rather is observed only in the imbalanced and not in the balanced training. We hypothesize that the class imbalance may adversely affect the speed at which individual features are learned and thus lead to non-monotonic training trajectories.

\subsection{Early Stopping and Overparameterization Benefit}

Second, we investigate model-wise generalization more closely and observe additional nuance emerge when considering early stopping, task complexity, and dataset size, i.e. when the number of classes and number of samples available per class is scaled up. We extend the balanced 4-way classification task to a 15-way classification taking into account a more complete set of Standard Model background processes.  As in the previous cases, model-wise double descent is observed when training long past the interpolation threshold. In both cases, the high-capacity models do not improve over the classical regime when comparing the final test-time loss.

\begin{figure}[H]
  \centering
  \includegraphics[width=0.45\textwidth]{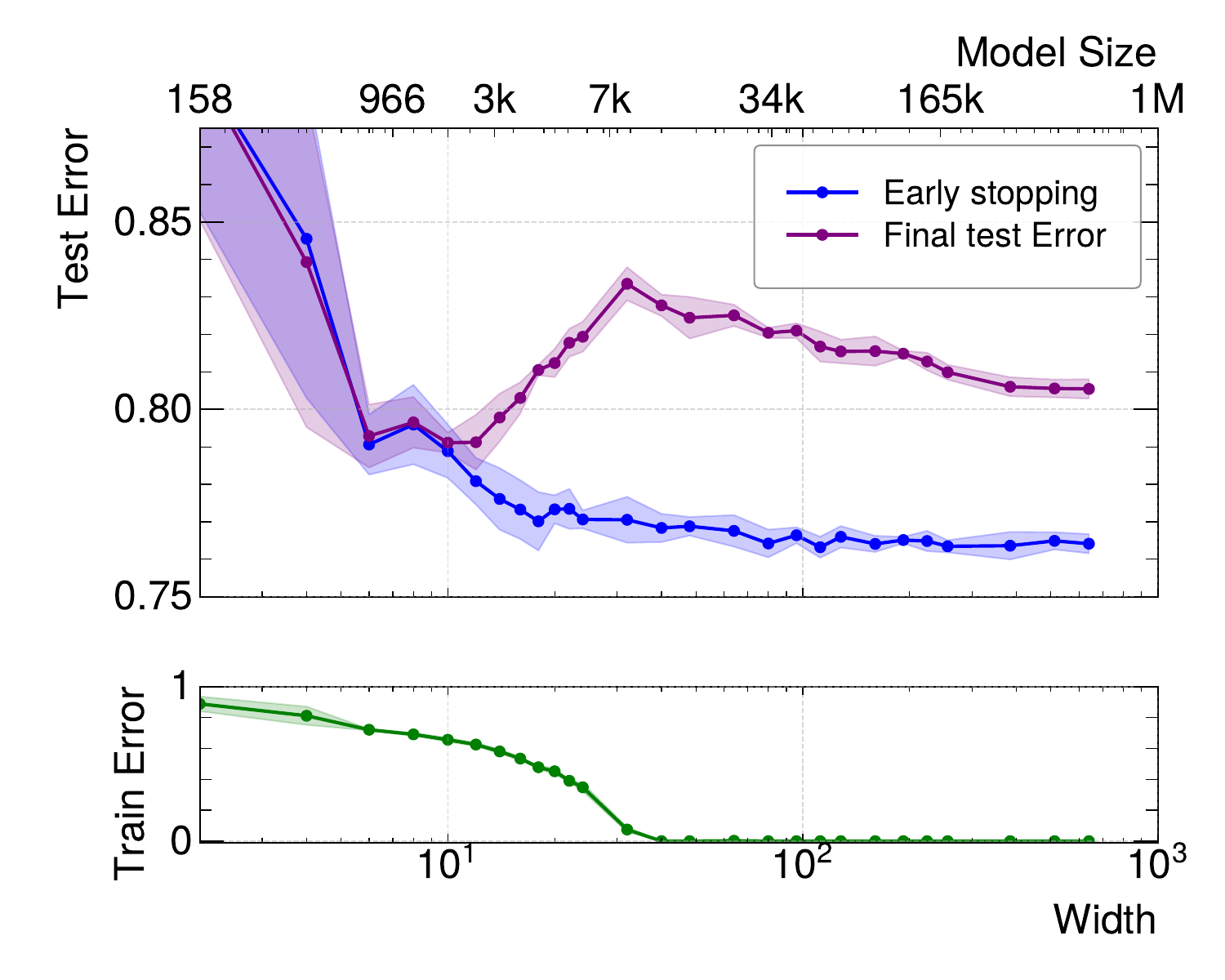}
  \includegraphics[width=0.45\textwidth]{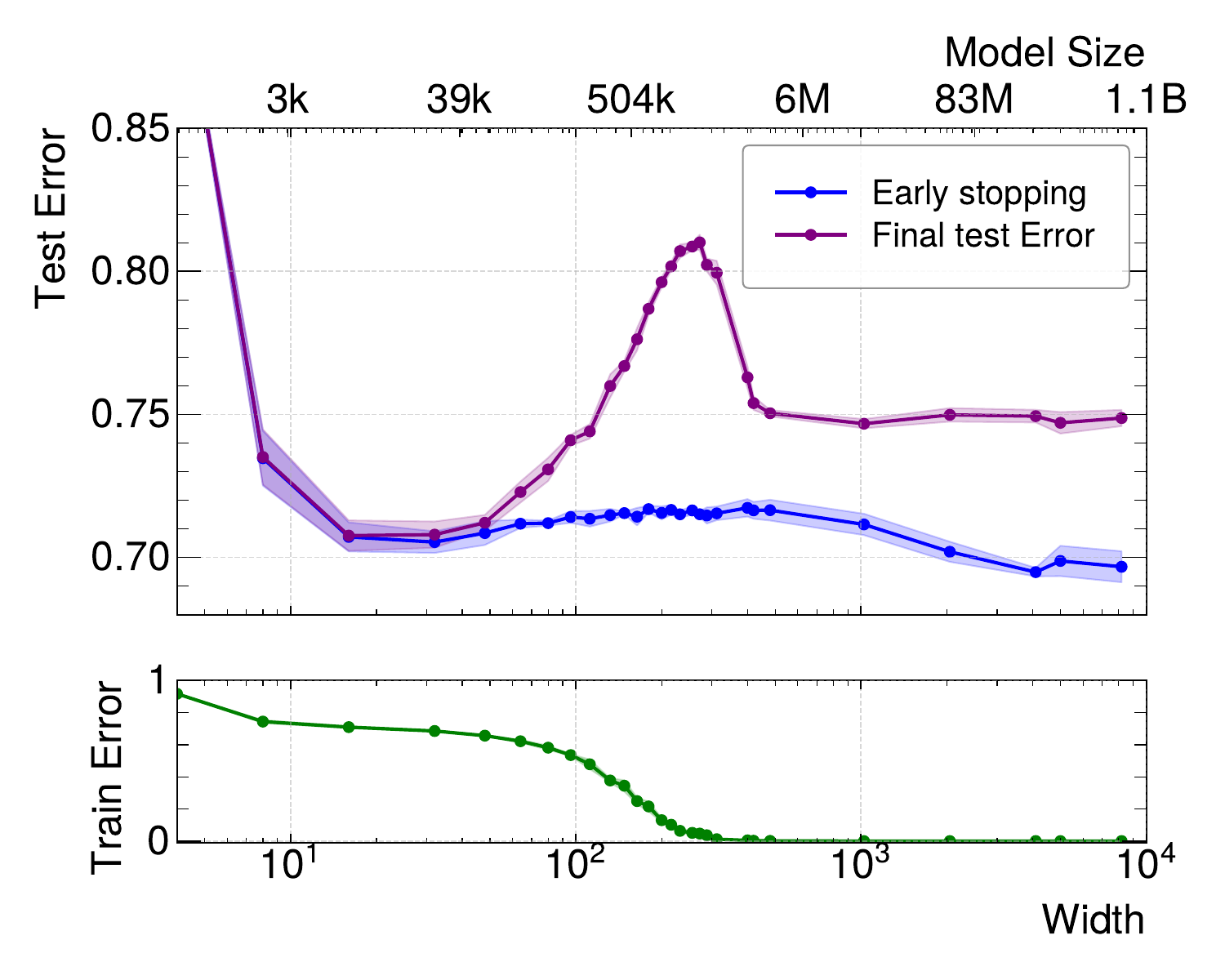}
  \caption{Model-wise double descent for a 3-layer MLP trained on 3k events (left) and a 16-layer MLP trained on 150k events (right). overparameterised models outperform underparameterised ones.}
  \label{fig:200}
\end{figure}

The picture changes when considering ``early-stopping'' in the sense of taking the epoch checkpoint with the best test-set performance. As shown in Fig.~\ref{fig:200}, for small datasets the generalization penalty at the interpolation peak disappears completely, and a monotonic improvement in test-time performance is observed for heavily overparameterized models. When increasing dataset size the monotonous behavior stops and a temporary increase in generalization error appears around the interpolation threshold, i.e. the double descent pattern emerges. However, unlike the previous examples, the high-capacity models clearly outperform models in the classical regime. In short, the recently observed benefit of training models with far more parameters than data samples can also be observed in particle physics data: a 1B parameter model outperforms classical models on a comparatively modest dataset.

\section{Conclusions}

We investigated the behavior of under- and overparameterized models on high-energy physics data. We show that neural networks exhibit a rich range of behaviors: depending on context, both model- \emph{and} epoch-wise double descent is observed both with and without early stopping. We identified cases where overparameterized models significantly outperform models from the classical regime. As the field moves towards higher-capacity models it is essential to further understand these phenomena. 

\section{Acknowledgements}

The authors thank Nicole Hartman for valuable discussions and a careful read of the manuscript. LH is supported by the Excellence Cluster ORIGINS, which is funded by the Deutsche Forschungsgemeinschaft (DFG, German Research Foundation) under Germany’s Excellence Strategy - EXC-2094-390783311.

\bibliographystyle{unsrtnat}
\bibliography{./main.bib}
\appendix


\end{document}